# Characterization and compression of dissipative-soliton-resonance pulses in fiber lasers

**Daojing Li,[1] Lei Li,[2] Junyu Zhou,[3] Luming Zhao,[2, *] Dingyuan Tang,[2] and Deyuan Shen[1]**

[1]Department of optical science and engineering, Fudan University, Shanghai 200433, China

[2]Jiangsu Key Laboratory of Advanced Laser Materials and Devices, School of Physics and Electronic Engineering,

Jiangsu Normal University, Xuzhou, Jiangsu, 221116, China

[3]School of Electrical and Electronic Engineering, University of Adelaide, SA 5005, Australia

[*]Corresponding author. zhaoluming@jsnu.edu.cn

**ABSTRACT:** We report numerical and experimental studies of dissipative-soliton-resonance (DSR) in a fiber laser with a nonlinear optical loop mirror. The DSR pulse presents temporally a flat-top profile and a clamped peak power. Its spectrum has a rectangle profile with characteristic steep edges. It shows a unique behavior as pulse energy increases: The rectangle part of the spectrum is unchanged while the newly emerging spectrum sits on the center part and forms a peak. Experimental observations match well with the numerical results. Moreover, compression of the DSR pulses is both numerically and experimentally demonstrated for the first time. An experimentally obtained DSR pulse of 63 ps duration is compressed down to 760 fs, with low-intensity pedestals using a grating pair. Before being compressed to its narrowest width, the pulse firstly evolves into a cat-ear profile, and the corresponding autocorrelation trace shows a crown shape, which distinguishes itself from properties of other solitons formed in fiber lasers.

## I.    INTRODUCTION

Passively mode-locked fiber lasers have been extensively investigated as a simple and reliable ultrafast pulse source that favors various scientific, industrial, and biomedical applications [1-2]. Solitons as the consequence of a natural balance between the anomalous dispersion and fiber nonlinear Kerr effect can be generated in the anomalous-dispersion mode-locked fiber lasers. In fiber lasers excessive nonlinear phase accumulation severely limits the pulse energy by causing pulse breaking [3]. To deal with this issue, two basic approaches have been carried out in fiber lasers. One is to minimize the nonlinear effect directly, including lowering the pulse peak power by a variety of methods, such as stretching the pulses with dispersion management [4], giant-chirp oscillators [5], divided pulse lasers [6]. The other is to seek new mechanisms that can tolerate more nonlinear phase accumulation, such as self-similariton [7] and dissipative soliton formation [8, 9]. These methods have successfully scaled up the pulse energy in standard fiber lasers to tens of nano-joules. Nonetheless, wave breaking caused by nonlinear phase shift still imposes a fundamental obstacle on the achievable pulse energy. To pursue higher pulse energies, methods to circumvent pulse breaking is highly desired.

Recently, a new soliton formation, dissipative-soliton-resonance, predicted by Chang et al. in the frame of complex cubic-quantic Ginzburg-Landau equation has attracted considerable attentions as it supports wave breaking free pulses [10,11]. Under the DSR generation, with increasing gain, the pulse remains its peak power, while it keeps broadening in the time domain. Pulse breaking is avoided in this case. Therefore, DSR shows great potential for extremely high pulse energy generation. Lots of efforts have been devoted to obtain DSR in mode-locked fiber lasers. DSR is demonstrated and characterized in many theoretical models [12-14]. Previously, we have also numerically studied the physical mechanism of how the DSR generation could prevent the pulse breaking in all-normal-dispersion regime [15]. We found that with strong enough peak-power-clamping induced by sinusoidal saturation absorption such as nonlinear polarization rotation (NPR) or nonlinear optical loop mirror (NOLM), the pulse spectrum could be confined within the cavity filter transmission window, and thus pulse breaking is avoided. Experimental studies of DSR have also been carried out in a variety of fiber laser settings [16-20]. Experimentally square-wave pulses without wave breaking are observed, which stretched from picoseconds to nanoseconds. However, up to now, main efforts are focused on pulse temporal properties, namely square shape and wave-breaking-free. The spectral properties of DSR pulses are less investigated. A detailed characterization of DSR spectrum will be of high interest.

As being continuously broadened with increasing pump, the DSR pulses are normally heavily chirped. Chirped pulses are possible to be compressed using linear dechirp devices. DSR compression has been mathematically discussed in [11]. The author showed that as the chirp is far from linear, the dechirped pulse is multi-peaked. Later, the author also indicated that near resonance point, solitons could be compressed to Fourier-transform-limited pulses along with low-intensity pedestals [21]. Experimentally, the compression of DSR pulses so far has not been reported. Therefore, although DSR has shown great potential for achieving large pulse energy and nanosecond pulses, its practical application as an ultrafast light resource is hindered.

Here, we report an investigation of DSR generation in an all-normal-dispersion Yb-doped fiber laser (YDF) mode-locked by a NOLM. Experimental and numerical characterizations on both the pulse's temporal and spectral behaviors are presented. The experimental results match well with the simulations, confirming that the laser works in the resonance regime. Furthermore, we demonstrate the compression of DSR pulses for the very first time. Numerical result has shown that the DSR pulses could be dechirped with linear devices. Experimentally an DSR pulse of 63 ps duration is compressed down to 760 fs by a grating pair. During compression, the pulse shows distinctly a crown shape autocorrelation curve that distinguishes it from any other soliton formation. We believe that the demonstrated compressibility of the DSR pulses would significantly improve its usability as an ultrafast resource.

## II.  RESULTS

### a. Numerical Simulations

In [15], we had theoretically demonstrated DSR operation and suggested two key factors to achieve DSR in dissipative soliton fiber lasers: increasing the spectral filter bandwidth (BW) and introducing strong peak power clamping effect. However, there the saturable absorber was modeled by a simple transmission curve. To gain deeper insight of DSR operation in mode-locked fiber lasers with NOLMs, numerical simulation based on a more explicit model of NOLM was performed for the cavity illustrated in Fig. 1. Further details are illustrated in the METHODS. The NOLM was chosen to provide periodical absorption, as its transmission is practically easy to manipulate. It relies on nonlinear interference of the counter-propagating fields so its saturation power is inversely proportional to the product of splitting ratio and loop length. To obtain low saturation power, we chose an 80:20 coupler and a total 5-meter-long single mode fiber (SMF) forming the loop.

Numerical results show that under the current laser configuration, stable self-starting DSR solutions can be obtained. Characteristic results for the laser operation with increasing gain are plotted in Fig.2. With the increase of gain saturation energy $E_{sat}$, the pulse peak power increases slightly at first, then reaches its maximum, as shown in Fig. 2(a). The dotted lines are the corresponding frequency chirps of the pulse. Increasing the gain, the pulse develops into a flat-top profile. The newly generated energy will locate at the center of the pulse and the pulse begins to expand. The pulse energy and width increase linearly with the gain saturation energy $E_{sat}$ [Fig. 2(b)]. During the broadening, the pulse edges maintain themselves and show the same large linear chirp, whereas a low linear chirp prevails across the extended pulse plateau. The temporal and chirp profiles agree well with DSR results predicted by other models [10-14].

For the pulse spectrum, it exhibits a rectangle profile with steep edges, just as regular dissipative solitons formed in normal-dispersion fiber lasers [Fig. 2(c)]. However, it shows a unique behavior with increasing energy: Its rectangle part is unchanged while the newly generated energy locates on the center part of the spectrum. As the pulse peak power keeps constant, the spectral edge-to-edge BW remains the same. This is well expected since spectral broadening in fibers is primarily due to the self-phase modulation. The accumulation nonlinear phase shift is closely linked to the peak power, given by $\Phi_{NL} = \gamma PL$, where $\gamma$ is the nonlinearity coefficient, $P$ is the pulse power and $L$ is the propagation distance. The spectral behavior agrees well with the temporal and chirp profiles: The pulse edges have large chirp corresponding to the rectangle part of spectrum and maintain themselves during energy scaling; the central spectral part has low chirp and concentrates to form a peak near the central wavelength. As a result, with increasing gain, the spectral 3-dB BW slowly decreased, whereas the edge-to-edge BW keeps a constant [Fig. 2(d)].

### b. Experimental Results

Following the simulations, the laser of Fig. 1 was built. With appropriate settings of the PCs, mode-locking can be self-started, and can be maintained with the pump power being decreased to 200 mW. The results are summarized in Fig. 3(a-c). The optical spectrum [Fig. 3(a)] has steep spectral edges with a 3-db BW of 4.5 nm, indicating coherent mode-locking in normal-dispersion regime. Figure 3b shows the single pulse oscilloscope trace with a measured pulse width of 63 ps. The output power is 40 mW, corresponding to a pulse energy of 2.96 nJ at 13.4 MHz repetition rate. The autocorrelation trace [Fig. 3(c)] shows a triangle shape, as expected for the flat-top DSR pulse. The calculated time-bandwidth product (TBP) is 85, suggesting that the pulse is largely chirped. The radio frequency (RF) spectrum shows a high contrast of more than 70 dB, indicating low-amplitude fluctuations [inset of Fig. 3(c)]. The resolution bandwidth (RBW) of the RF spectrum analyzer is set at 10 Hz. Single-pulsing was confirmed by the combined measurement of the autocorrelation and oscilloscope traces.

The pump power was then slowly increased to its maximum. No multiple pulses were observed. The detailed pulse features with the increasing pump are plotted in Fig. 3(d-h). The optical spectrum slowly shifts toward short wavelength due to the reduced gain re-absorption with increasing pump [Fig. 3(d)]. For better comparison, the central wavelengths at different pump level are artificially shifted together in Fig. 3(e). One can see clearly that the rectangle part of spectrum almost keeps the same during energy scaling and the energy concentrates and forms a peak at the central wavelength. Figure 3(f) provides the measured variation of spectral edge-to-edge BW and 3-dB BW versus the pump power. The pulse oscilloscope trace presents a flat-top profile [Fig. 3(g)]. The output power and pulse width increase linearly with the pump power [Fig. 3(h)], also indicating a clamped peak power.

Figure 2 and Fig. 3 show good matching between the simulations and the experiments, revealing that except for the wave-breaking-free flat-top temporal profile, the DSR in normal-dispersion regime also shows distinct spectral properties: It has a rectangle spectrum with characteristic steep edges of dissipative solitons in normal-dispersion fiber laser. However, the rectangle part of spectrum keeps unchanged with the increasing pump. The newly generated part will sit on the spectrum and form a peak around the central wavelength. Its edge-to-edge BW keeps constant while its 3-dB BW decreases slowly.

c. **Compression of DSR pulses**

The existence of the extended chirp ensures significant room for pulse compression. To gain broader insight of its physical interest for ultrafast, we then both numerically and experimentally investigated the compression of the laser output. Numerical compression with a linear device was first performed, which is achieved by pulse propagation through anomalous $\beta_2$. Figure 4 summarizes the typical compressed pulses (a-c) and the corresponding autocorrelation traces (d-f) of the laser output ($E_{sat}$ = 0.8 nJ) versus dispersion compensation increase. Interestingly, the DSR pulse exhibits a unique compression dynamics different from any other solitons. As illustrated in Fig. 2a, the DSR pulse comprises two different chirps: large linear chirps at the both edges and a very low chirp across the center regime. Due to the different chirps, the edges with much larger chirp are compressed more quickly than the central part. Consequently, while increasing the dispersion compensation, the compressed pulse gradually develops into a cat-ear profile with high spikes at both edges and corresponds to a crown shape autocorrelation trace [$\beta_2$ = -2.4 ps$^2$, Fig. 4(a,b)]. Further increasing the dispersion the pulse continues to be compressed and shows plentiful oscillatory patterns. Figure 4(e) shows the compressed pulse and its frequency chirp when the dispersion compensation is increased to $\beta_2$ = -3.8 ps$^2$. As one can see, the output is compressed to an ultrafast pulse close to a Gaussian shape with small pedestals. It shows chirp-free across the main pulse indicating the pulse is fully dechirped. The resulting TBP is 0.88. Pulses with longer duration show similar compression dynamics except that larger dispersion compensation is needed for full compression.

We then experimentally compressed the laser output utilizing a grating pair. The output with the smallest duration of 63 ps [Fig. 3(c)] was first investigated. The measured autocorrelation traces of the compressed pulses with increasing dispersion compensation are presented in Fig. 4(g-i). The experimental results again agree well with numerical simulations. The pulse exhibits the characteristic crown shape autocorrelation trace during compression, further confirming the DSR operation. The pulse can be dechirped to 760 fs duration [Fig. 4(i)], assuming a Gaussian shape, with a compression ratio of 83. The resulting TBP is 1.02, close to the numerical simulation result. Large dispersion compensation amount (-6.48 ps$^2$) is needed to fully dechirp the pulse, which is almost the maximum dispersion our grating pair can supply. As for pulses with larger duration, more dispersion compensation is needed for full compression. Meanwhile the crown shape autocorrelation curves are always observed during compression. Large dispersion compensation is a challenge for a grating pair and the compressed pulses suffer from unwanted pedestals. For better compression, chirped fiber grating [22] or other more sophisticated dispersive lines can be referred to [11,21]. The results demonstrate that the DSR pulse can be dechirped to a near transform-limited pulse by a linear device.

In conclusion, we have studied DSR operation in an YDF laser with a NOLM. The DSR pulses exhibit a flat-top profile. Spectrally, they have a rectangle spectrum with characteristic steep edges. However distinctly, with increasing energy, the rectangle part of the spectrum is unchanged while the newly generated spectrum sits upon the center part and forms a peak. Moreover, compression of a DSR pulse is demonstrated for the first time. The pulse firstly evolves into a cat-ear profile with a unique crown shape autocorrelation curve, and finally can be compressed from original 63 ps down to a 760 fs pulse with small pedestals, by a grating pair. Experimental and numerical results match well. The DSR has already been shown highly appealing for delivering unlimited pulse energy. The demonstrated compressibility of the DSR pulse would considerably improve its usability as an ultrafast resource.

## III. METHODS

Numerical simulations are based on the complex nonlinear Ginzbug-Laudan equation:

$$\frac{\partial u}{\partial z} = -\frac{i\beta_2}{2}\frac{\partial^2 u}{\partial t^2} + i\gamma |u|^2 u + \frac{g}{2}u + \frac{g}{2\Omega_g^2}\frac{\partial^2 u}{\partial t^2}.$$

Here, $u$ is the slowly varying pulse amplitude envelop, $z$ is the propagation distance, and $t$ is the local time. $\beta_2$ and $\gamma$ represent the second-order dispersion and the fiber nonlinearity, respectively. All the fibers are assumed with the same dispersion $\beta_2 = 22$ ps$^2$/km, and nonlinearity $\gamma = 5.8$ (W·km)$^{-1}$. The saturable gain of the Yb fiber is given by

$$g = \frac{G_0}{1 + \int |u|^2 dt / E_{sat}}.$$

where $G_0 = 6.9$ m$^{-1}$ is the small-signal-gain, corresponding to 30 dB/m, $E_{sat}$ is the saturation energy, $\Omega_g$ is the gain BW, assumed as 40 nm. The cavity parameters are set as to match up with their experimental values. A more explicitly model of NOLM was adapted. Briefly, the light field is divided when it goes through the NOLM coupler. Then two fields propagate individually along their respective directions before they are recombined in the NOLM coupler, where the splitting ratio is applied again. The numerical model is solved with the split-step Fourier method and the simulation starts with white noise. The same stable solutions could be reached from different initial conditions.

Experimentally, the laser is counter-pumped by a 976 nm laser diode through a wavelength division multiplexer (WDM). No spectral filter is inserted in the cavity. The YDF (YB406, Coractive) is 42 cm long. A 30% coupler is used to output the light and an isolator is adopted to ensure unidirectional operation. Two polarization controllers (PCs) in the cavity and the loop permit optimization of the cavity birefringence and loss. The total cavity length is 15.7 m. The laser is monitored by a photodetector-oscilloscope combination with a BW of 45 GHz and an intensity autocorrelator with a scan range of 150 ps.

**Acknowledgements**

This work was supported in part by the National Natural Science Foundation of China under Grant 61275109, 61177045, 11274144, and 61405079, in part by the Priority Academic Program Development of Jiangsu higher education institutions (PAPD), in part by the Jiangsu Province Science Foundation (BK20140231).

**Competing financial interests:** The authors declare no competing financial interests

# Figures

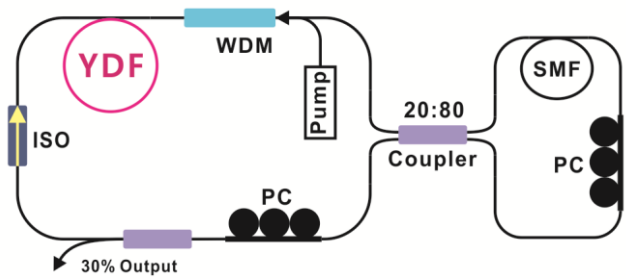

Fig. 1. Schematic of the laser setup. WDM, wavelength division multiplexer, ISO, isolator, PC, polarization controller, YDF, ytterbium-doped fiber, SMF, single-mode fiber.

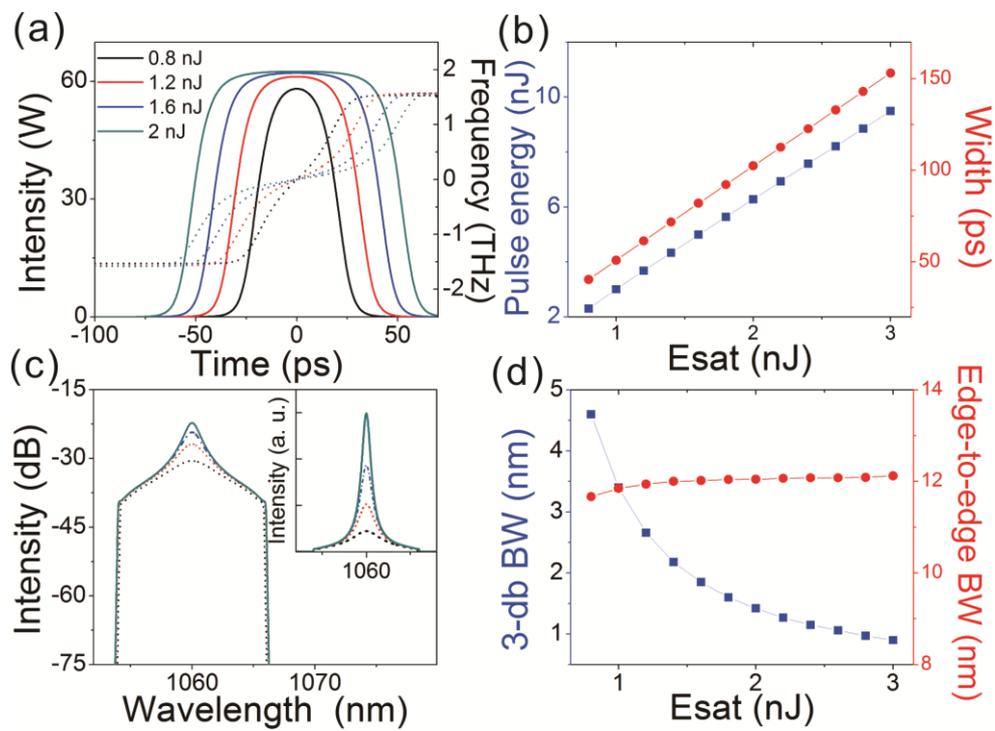

Fig. 2. Numerical results with increasing gain saturation energy $E_{sat}$. (a) pulse temporal profiles (solid) and frequency chirps (dashed), (b) variation of pulse energy (blue square), width (red dot), (c) optical spectra (inset: spectra in linear scale), (d) variation of spectral 3-db BW (blue square) and edge-to-edge BW (red dot).

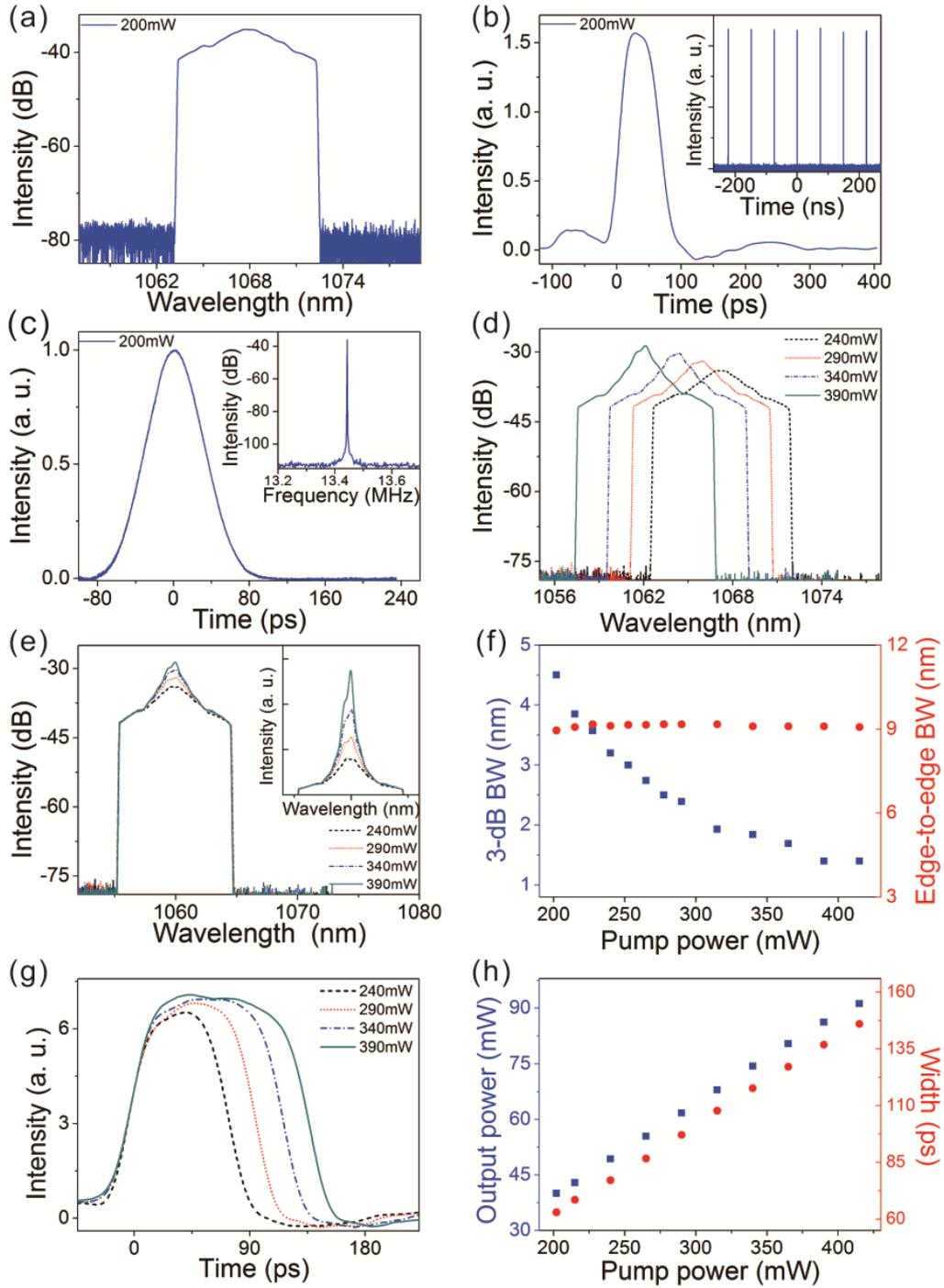

Fig. 3. Experimental results of the laser output. (a) Optical spectrum, (b) single pulse shape (inset: pulse train), (c) autocorrelation trace (inset: RF spectrum), under pump power $P_p$ = 200 mW. (d) measured spectra, (e) wavelength artificially shifted spectra (inset: spectra in linear scale), (f) variation of spectral 3-dB BW (blue square) and edge-to-edge BW (red dot), (g) oscilloscope traces, (h) variation of output power (blue square) and pulse width (red dot), with increasing pump power.

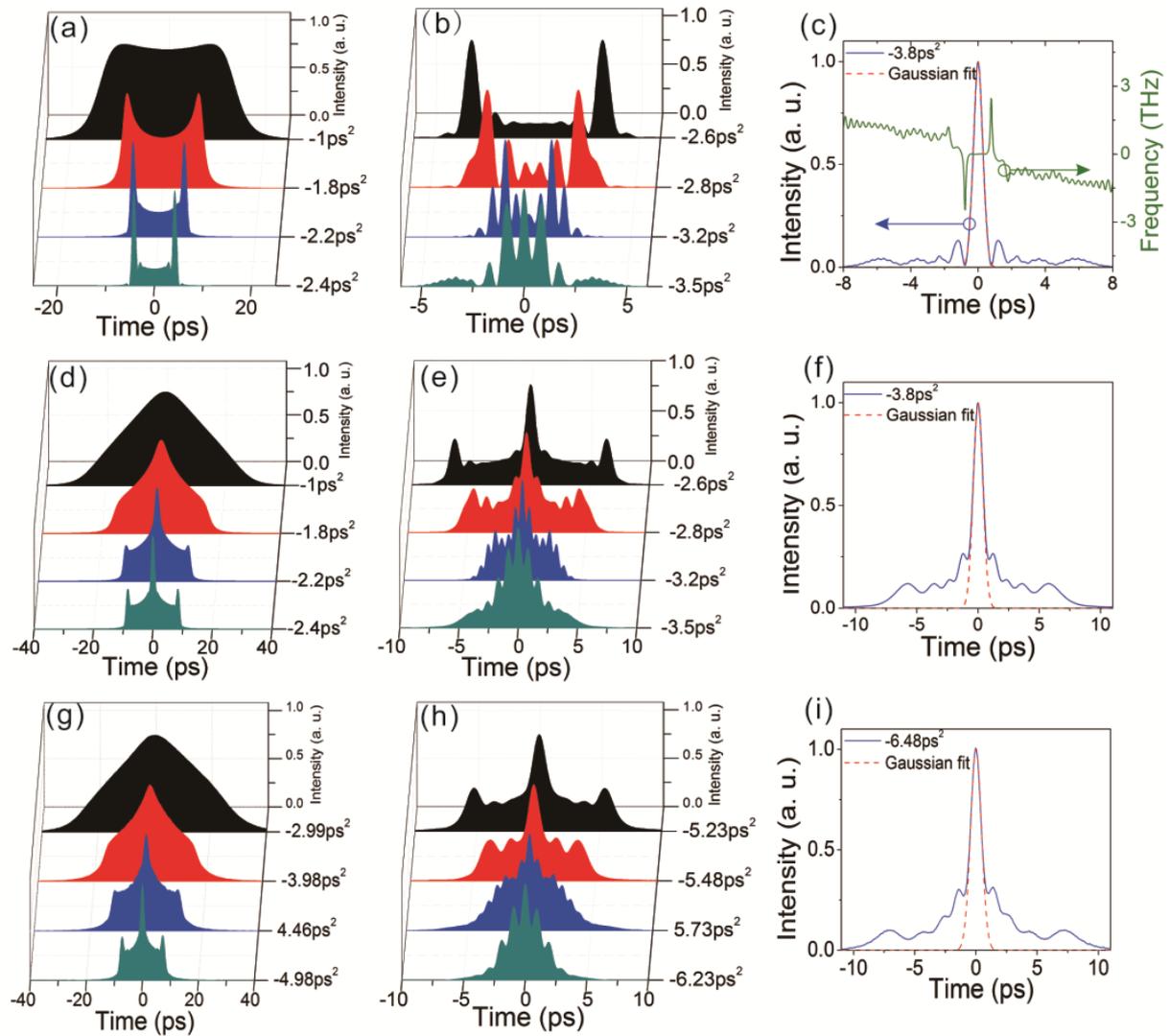

Fig. 4. (a), (b) Numerically simulated compressed pulse profiles versus dispersion compensation, (c) chirp-free pulse profile after compression (blue solid), Gaussian fit (red dashed) and frequency chirp (green solid). (d, e, f) The autocorrelation traces of (a, b, c). (g, h) Experimentally measured autocorrelation traces versus dispersion compensation, (i) autocorrelation trace of compressed output with the shortest duration (blue solid) and Gaussian fit (red dashed).